\def\Ol{\Omega_\Lambda}
\def\Om{\Omega_m}

\documentclass[structabstract]{aa}

\usepackage{txfonts}
\usepackage{graphicx}
\usepackage{epsf}

\begin{document}

\title{Testing for evolution in scaling relations of galaxy clusters: Cross analysis between X-ray and SZ observations}

\author{
L.D. Ferramacho\inst{1,2} \and A. Blanchard\inst{1,2}
}

\institute{$^1$ Universit\'e de Toulouse; UPS-OMP; IRAP; Toulouse, France\\
 $^2$ CNRS; IRAP; 14, avenue Edouard Belin, F-31400 Toulouse, France\\
\email{luis.ferramacho@ast.obs-mip.fr \& alain.blanchard@ast.obs-mip.fr}  
}

\offprints{luis.ferramacho@ast.obs-mip.fr}

\abstract{}
{ We present predicted Sunyaev-Zeldovich (SZ) properties of known X-ray clusters of galaxies for which gas temperature measurements are available. The reference sample was compiled from the BAX database for X-ray clusters. }
{The Sunyaev-Zeldovich signal is predicted according to two different scaling laws for the mass-temperature relation in clusters: a standard relation and an evolving relation that reproduces well the evolution of the X-ray temperature distribution
 function in a concordance cosmology. Using a Markov Chain Mote Carlo (MCMC) analysis we examine the values of the recovered parameters and their uncertainties.}
{The evolving case can be clearly distinguished from the non-evolving case, showing that SZ measurements will indeed be efficient in constraining the 
thermal history of the intra-cluster gas. However, significant bias appears in the measured values of the evolution parameter for high SZ threshold owing to selection effects.}
{}


\authorrunning{Ferramacho \& Blanchard}
\titlerunning{Scaling relations of galaxy clusters from X-ray and SZ observations}

\maketitle

\section{Introduction}


Galaxy clusters are the largest gravitationally bound objects in the Universe and can therefore provide important information on both the evolution of large-scale structures and the astrophysics of collapsed gas and dark matter. In particular, the abundance of massive clusters has been shown to have the potential to provide strong constraints on the cosmological parameters, because it depends exponentially on the cosmological growth factor (\cite{OB}; \cite{BB}). Several authors have used cluster abundance measured through X-ray observations to derive cosmological constraints, but the results do not always agree (\cite{bsbl}; \cite{H04}; \cite{vikhlinin}; \cite{delsart}). This discrepancy arises because it is difficult to find a reliable estimator for the total mass of a cluster, since this quantity is not directly observable. Typically, the mass of a cluster is estimated with an assumption on the global physics, such as a  hydrostatic equilibrium of the hot gas in the dark matter potential well. This provides a way to connect the total mass at a scalable radius to a given observable such as the X-ray temperature. These indirect measurements together with the results of numerical simulations (\cite{Rasia06}; \cite{Kay07}) indicate that at least for local clusters the relation between mass and X-ray temperature agree quite well (despite a non-negligible dispersion) with what one would expect from simple scaling arguments based on the gravitational heating of the gas, with a possible uncertainty in its calibration. However, \cite{vauclair} modeled the abundance of X-ray galaxy clusters at high redshifts ($z \approx 0.5$) and showed that for observed cluster counts to be compatible with a standard cosmology dominated by dark energy one would have to consider a significant amount of non-standard evolution with redshift in the $M-T$ relation. A similar result was obtained recently by Delsart et al. (2010). Indeed, it has become generally accepted that non-gravitational heating of the gas is significant and can explain the observed departure from the $L-T$ scaling relation at z=0 (\cite{Voit05}). This complex source of heating could also induce an evolution in the $M-T$ relation (\cite{Nath11}), which may remain undetected even if distant clusters are already in hydrostatic equilibrium. Within this reasoning, one can consider that a given scaling law can still hold true at higher redshifts if a non-standard evolution factor is introduced. This factor could be constrained with good observational data up to high redshifts. Of particular interest is the cross-analysis between measurements of the Sunyaev-Zeldovich effect and X-ray temperature, because both data probe the physics of the hot gas and are related to the overall cluster mass through different scaling relations.  

The thermal Sunyaev-Zeldovich effect (\cite{SZ}) has become a very promising way to observe and detect new galaxy clusters. This effect corresponds to a shift observed in the Cosmic Microwave Background (CMB) frequency spectrum, which is caused by the inverse Compton scattering of the cold CMB photons by the high-energy electrons that compose the intramedium gas of galaxy clusters. The associated flux decrement can be analytically estimated and characterized by the central parameter $y_0$ or the total integrated flux $Y$. This provides a different way to investigate cluster abundances for cosmological constraints (\cite{Barbosa}; \cite{Aghanim97}). The SZ signal was already measured for some clusters and used to study scaling relations between SZ and X-ray observed clusters (\cite{morandi}). However, to better understand and expand these results, it is important to perform a full study of the practical implications of assuming different flux limits for the samples as well as having larger samples and smaller observational errors when deriving cosmological results.     

The purpose of the present paper is then to investigate the capability of present and future experiments in constraining a possible evolution with redshift of the $Y-T$ and $M-T$ scaling relations, identifying at the same time the eventual presence of bias owing to selection effects or other causes. For this we simulate the observed SZ integrated fluxes for clusters with known X-ray temperatures in two distinct fiducial models with different selection criteria.  

The paper is organized as follows. In Sect. 2 we briefly present the theory behind the scaling laws for galaxy clusters and a model for their evolution. In Sect. 3 we present the cluster sample and its simulated SZ observations for different observational limits. The results on cosmological constraints, normalization and redshift evolution are presented in Sect. 4. Finally, we present our conclusions in Sect. 5.         
     
\section{Scaling relations for $M-T$, $Y-M$ and $Y-T$}

To test the effect of combined measurements on the SZ integrated flux $Y$ and the X-ray emission-weighted temperature we start by considering a general form for the standard scaling law between the cluster's mass and temperature. The departure from the  expected evolution for pure gravitational heating is taken into account by introducing an additional term of $(1+z)^{\alpha}$, $\alpha$ being an unknown parameter. One can write

\begin{equation}
T=A_{TM}M_{15}^{2/3}\left(\frac{\Omega_M \Delta_{M}}{178}\right)^{1/3} h^{2/3}(1+z)^{1+\alpha}\;\rm keV \;,
\label{MTv}
\end{equation}  
$\Delta_M$ being the enclosed overdensity inside a cluster compared to the mean background density of the universe, or equivalently, for the same physical radius:  
\begin{equation}
T=A_{TM}M_{15}^{2/3}\left(\frac{\Delta_{c} E^2(z)}{178}\right)^{1/3}h^{2/3}(1+z)^{\alpha}\; \rm keV\;,
\label{MTc}
\end{equation}    
$\Delta_c$ being the enclosed overdensity compared to the critical density.
We emphasize that both expressions are physically equivalent and will proceed using Eq. 1 hereafter. One can interpret these expressions through a physical dependence of the temperature on the contrast overdensity at a given radius, the total mass at this radius (expressed in units of 10$^{15}$M$_{\odot}$), and a normalization constant ($A_{TM}$), which depends on $\Delta$. At the virial radius, the overdensity is usually computed from the spherical top-hat model for gravitational collapse, and for an Einstein-de Sitter universe, one has $\Delta_{M,V}=\Delta_{c,V}=18\pi^2$. An approximation to the general form of this quantity was provided by \cite{BN}, which allows us to determine the virial overdensity 
for any flat $\Lambda$CDM model. In the above expressions, the dependence of the background or critical density with redshift is encoded in the $(1+z)$ and $E(z)$ factors, respectively, with $E^2(z)=\Omega_m(1+z)^3+\Omega_R(1+z)^2+\Omega_{\Lambda}$. 

\begin{figure*}
\begin{tabular}{c c}
\includegraphics[scale=0.45,angle=0]{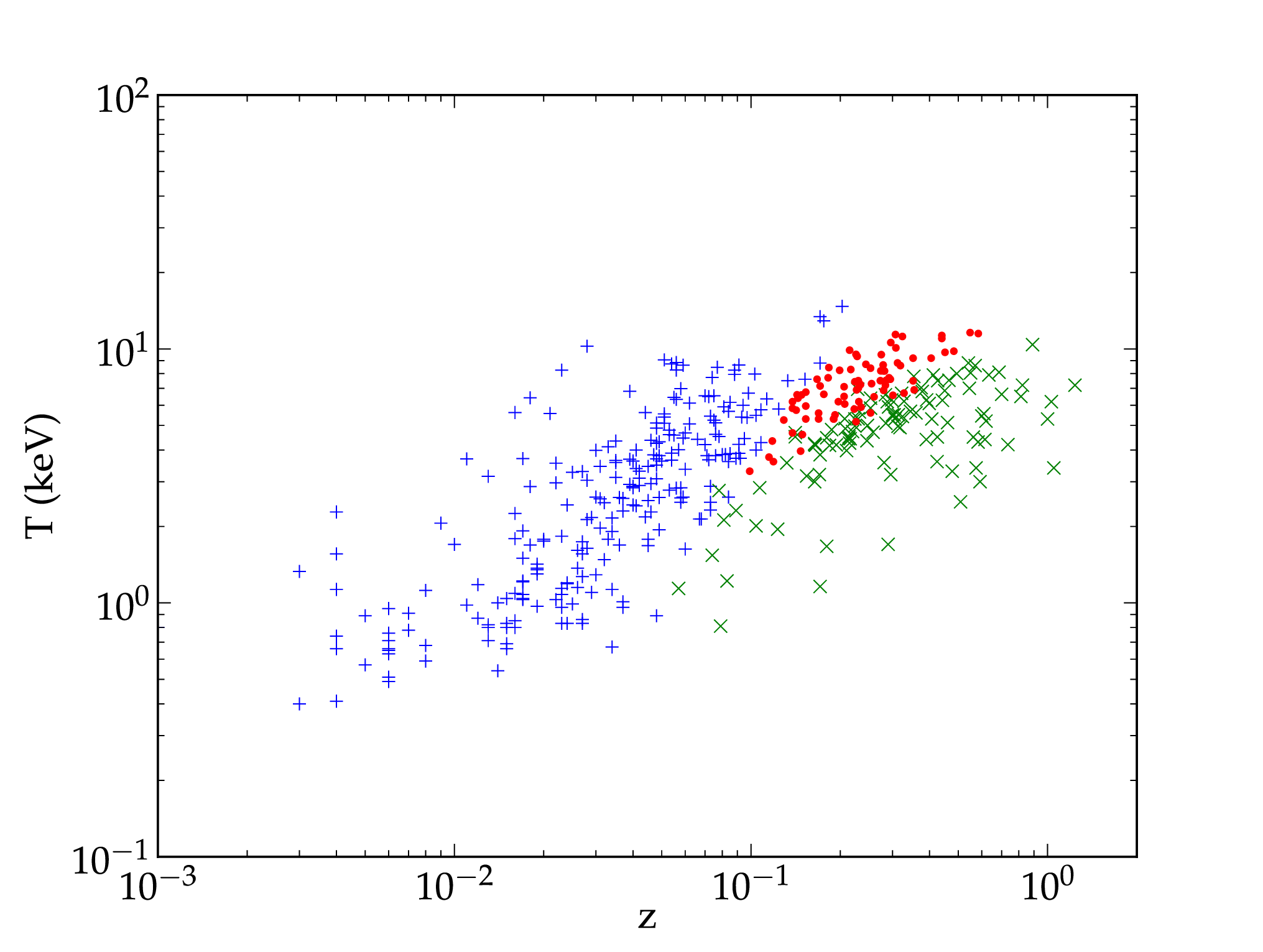}&
\includegraphics[scale=0.45,angle=0]{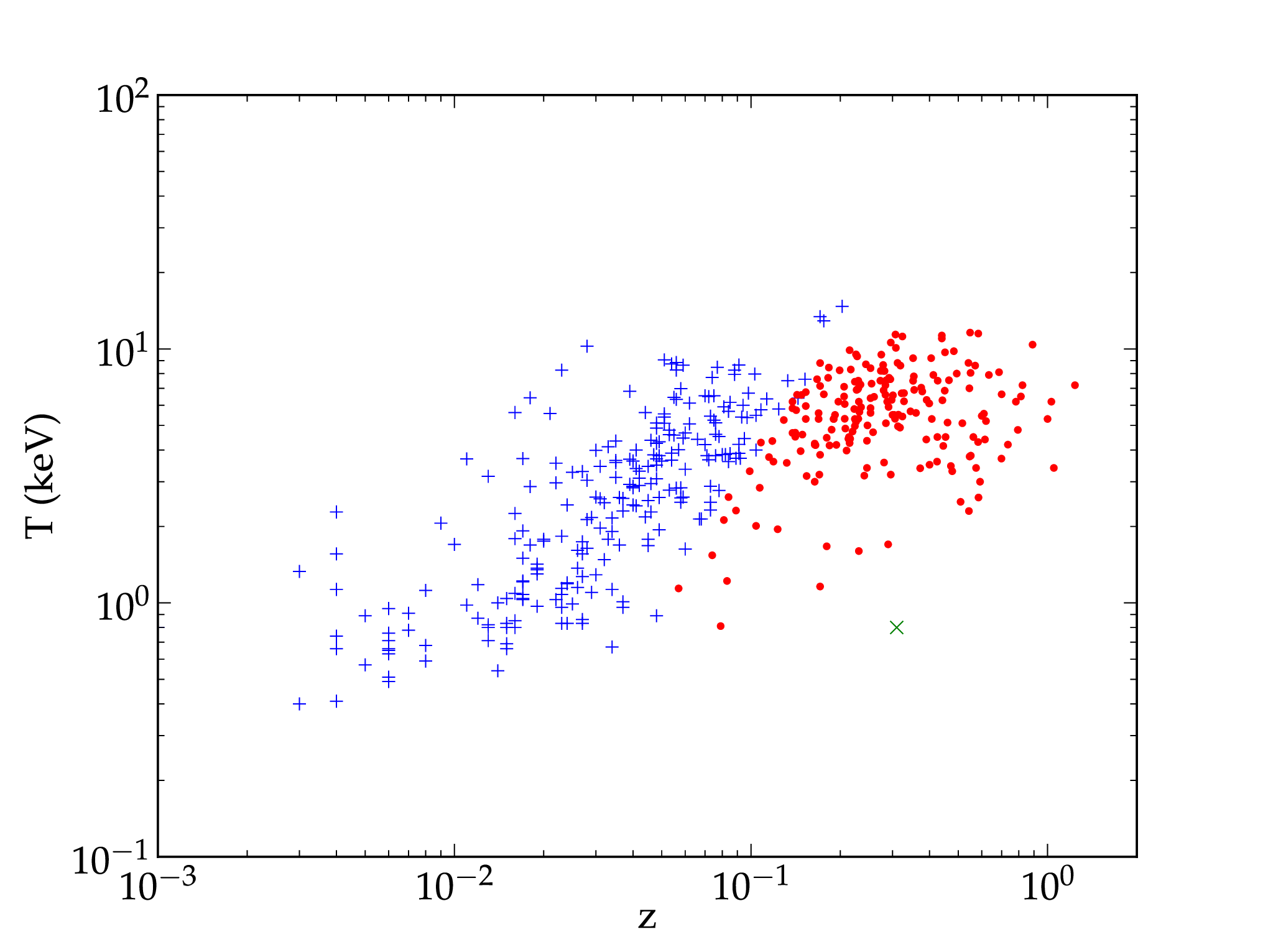}\\
\end{tabular}
 \caption{Distribution of temperature and redshift for the cluster sample used for this study and the clusters obtained after selection cuts for one Monte Carlo simulation. Blue plus signs represent clusters from the original sample that were excluded after the angular size cut; green crosses correspond to undetected clusters assuming a detection threshold of $1.10^{-3}$ arcmin$^2$ (\emph{left}) or $1.10^{-5}$ arcmin$^2$ (right). Red dots represent detected clusters under the same detection thresholds. This figure remains essentially unchanged in all simulations performed in this work.}
 \label{fig:z_T}
\end{figure*}

The above $M-T$ scaling relation (Eq. 1) can be used to derive a consistent expected relation between the integrated SZ flux $Y$ and $T$. The thermal SZ effect consists of a deformation of the CMB spectrum, which can be quantified by the flux (or Compton) $y$ parameter along the line of sight:       
\begin{equation}
y=\frac{k_B \sigma_T}{m_e c^2}\int n_e T_e dl\;.
\end{equation}         

Integrating $y$ over the solid angle $\Omega$ subtended by the cluster introduces the total $Y$ parameter:
\begin{equation}
Y =  \int_{\Omega}yd\Omega= \frac{k_B \sigma_T}{m_e c^2}\frac{1}{d_a^2}\int_0^\infty \int_S n_e T_e dA dl,  
\end{equation}         
where $d_a$ is the cosmological angular distance to the cluster and $S$ is the area of the cluster observed in the sky. $Y$ is an observable in SZ surveys and represents a unique way to directly measure the total thermal energy of the gas content in galaxy clusters. One can replace the integrated temperature by the mass-averaged gas temperature and write the remaining integral in terms of the total mass of the cluster. This allows to derive a general scaling law from the above expression:
                           
\begin{equation}
Y\propto T_g M_g d_a^{-2}=T_gf_gM_{tot}d_a^{-2}\;, 
\end{equation}  
$T_g$ being the mass-averaged temperature of the gas.  The gas fraction at the virial radius of clusters  is usually expressed as a fraction of the universal baryonic fraction ($f_g=\Upsilon\frac{\Omega_b}{\Omega_m}$). Direct measurements of the gas fraction up to the virial radius are difficult to achieve, but numerical simulations seem to agree with a value of $\Upsilon$ between 0.7 and 0.8  (\cite{DeBoni}). Here we assumed a fixed value of 0.8, consistent with a 15\% contribution of stars to the baryonic component (\cite{Roussel2001}). 
If the gas were completely isothermal in clusters, one would have $T_g = T_X$, with $T_X$ being the measured X-ray temperature and Eq.4 could then be used to probe the gas mass directly. However, it has been shown that clusters are not isothermal, with the temperature declining in the outer parts (\cite{vikhlinin05}). In this case, one can still assume that the gas temperature follows the scaling law but with a different normalization so that $T_g$=$\xi T_X$, the scaling assumption being that $\xi$ is a constant. This normalization can be estimated using the observed profiles for gas mass fractions (\cite{Sadat05}) and temperature (\cite{vikhlinin05}), together with the theoretical NFW profile for the dark matter halo. The calculation yields $\xi\sim  0.6$, a number that is subjected to a substantial uncertainty but that could be further constrained by observations. 

Using the mass-temperature relation from Eq.1, the above expression can be used to derive the full $Y-T_X$ scaling relation, which connects two directly observable quantities. Assuming that the SZ contribution from regions outside the virial radius can be neglected, we have: 
\begin{equation}
Y=1.816\cdot10^{-4} \xi A_{TM}^{-\frac{3}{2}} f_g\left(\frac{\Omega_M \Delta_V}{178}\right)^{-\frac{1}{2}}T_X^{\frac{5}{2}}\left(1+z\right)^{-\frac{3}{2}(1+\alpha)}h^{-1}d_a^{-2}\;.
\label{Y-T}
\end{equation}

If scaling holds, the existence of some evolution in the mass-temperature relation and the normalization between the gas and the X-ray emission-weighted temperature could be constrained using measurements of both $Y$ and $T_X$ for the same clusters. Below we address the constraining power of these measurements for known X-ray clusters by considering two fiducial cases: the standard scaling relation ($\alpha=0$) and a significant deviation caused by evolution; like \cite{vauclair} and Delsart et al. (2010) we adopted $\alpha=-1$ as a fiducial value.  As for the temperature normalization parameter $\xi$, we consider a constant fiducial value of 0.57.

\begin{figure}[t!]
\vspace{0.6cm}\includegraphics[width=0.48\textwidth]{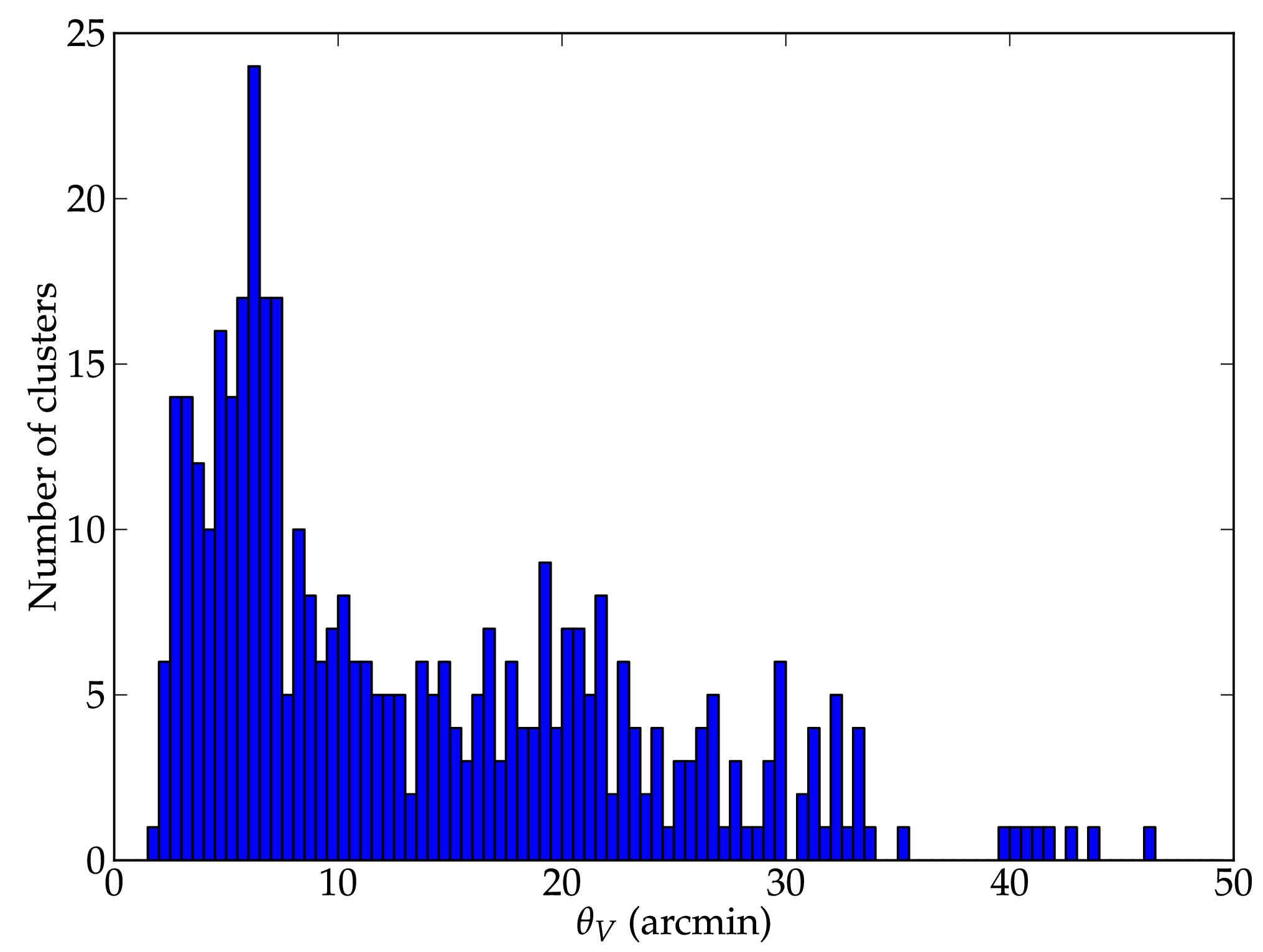}\vspace{-0.1cm}\\
\includegraphics[width=0.48\textwidth]{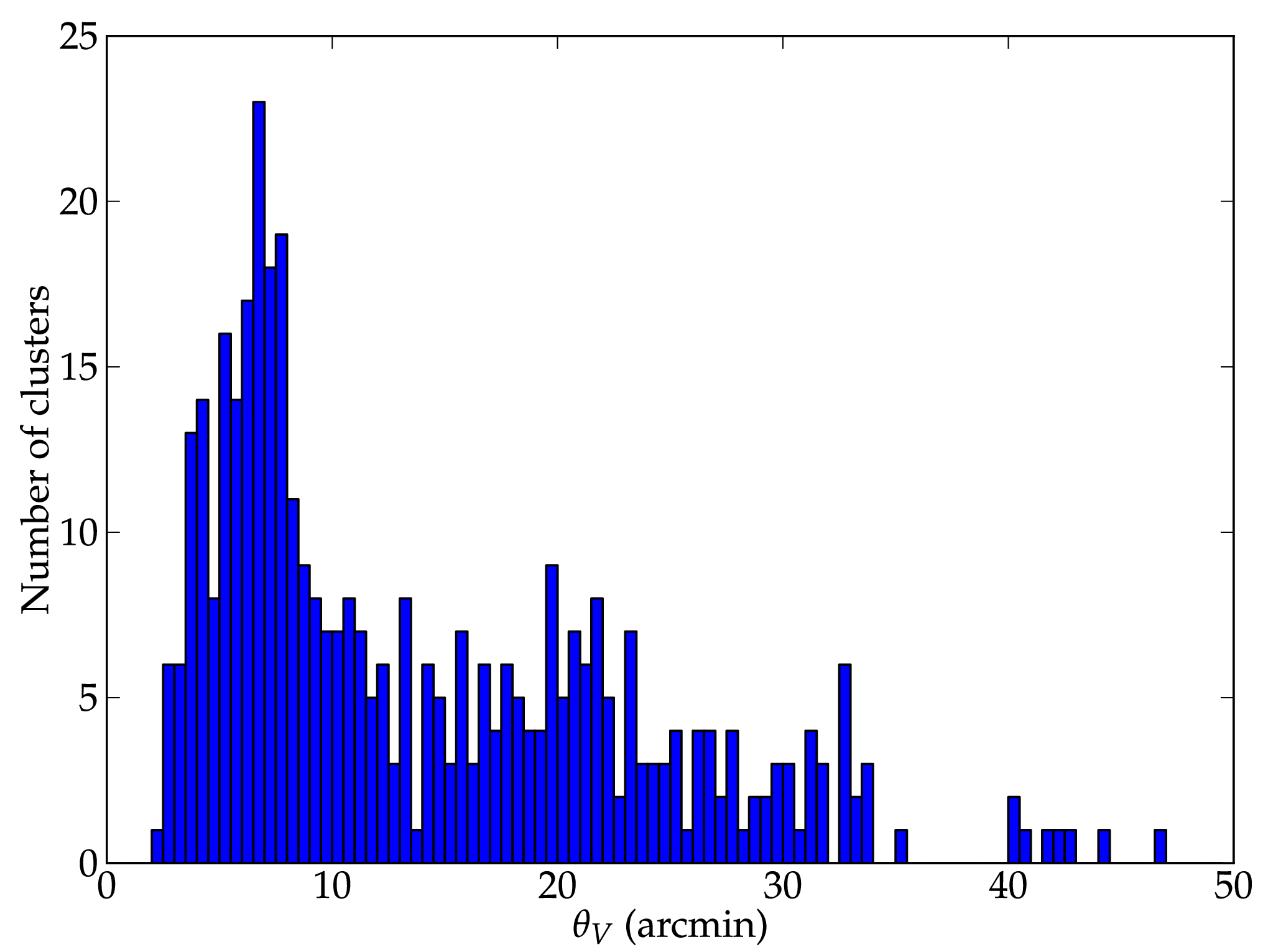}
\caption{ Histogram of the angular size of our cluster sample as expected from standard $M-T$ relations (\emph{top}) and accounting for a strong evolution in $M-T$ with $\alpha=-1$ (\emph{bottom}).}
\label{fig:size_dist}
\end{figure}

Unless otherwise noted, we performed our study using the best cosmological model derived through the COSMOMC\footnote{http://cosmologist.info/cosmomc/} package, using a combination of the WMAP 7-year data (\cite{wmap7}), SDSS LRG 7 (\cite{sdss7}) and the Constitution SNIa sample (\cite{const}). The best-fit model obtained is $\Omega_M=0.317$, $\Omega_{\Lambda}=0.682$, $\Omega_b=0.0499$, $h=0.671$, where $H_0 = 100\;h$ km s$^{-1}$ Mpc$^{-1}$. We assumed a fixed value of 8.21 keV for the normalization factor $A_{TM}$, calibrated through the analysis of the temperature function of observed clusters (\cite{delsart}).

\section{Simulated SZ observations of real clusters in two evolution scenarios}

\begin{figure}[t!]
\vspace{0.6cm}\includegraphics[width=0.48\textwidth]{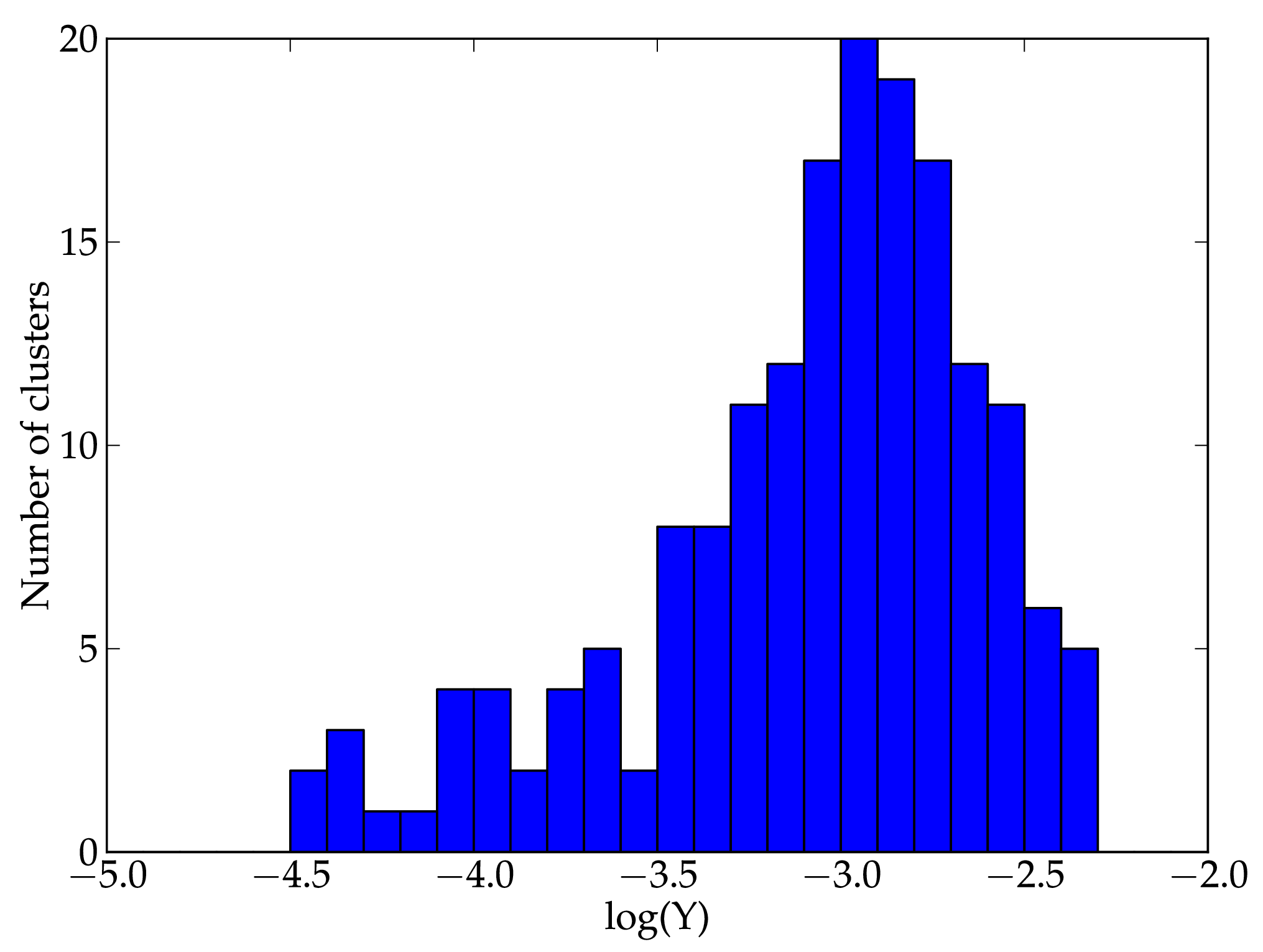}\vspace{-0.1cm}\\
\includegraphics[width=0.48\textwidth]{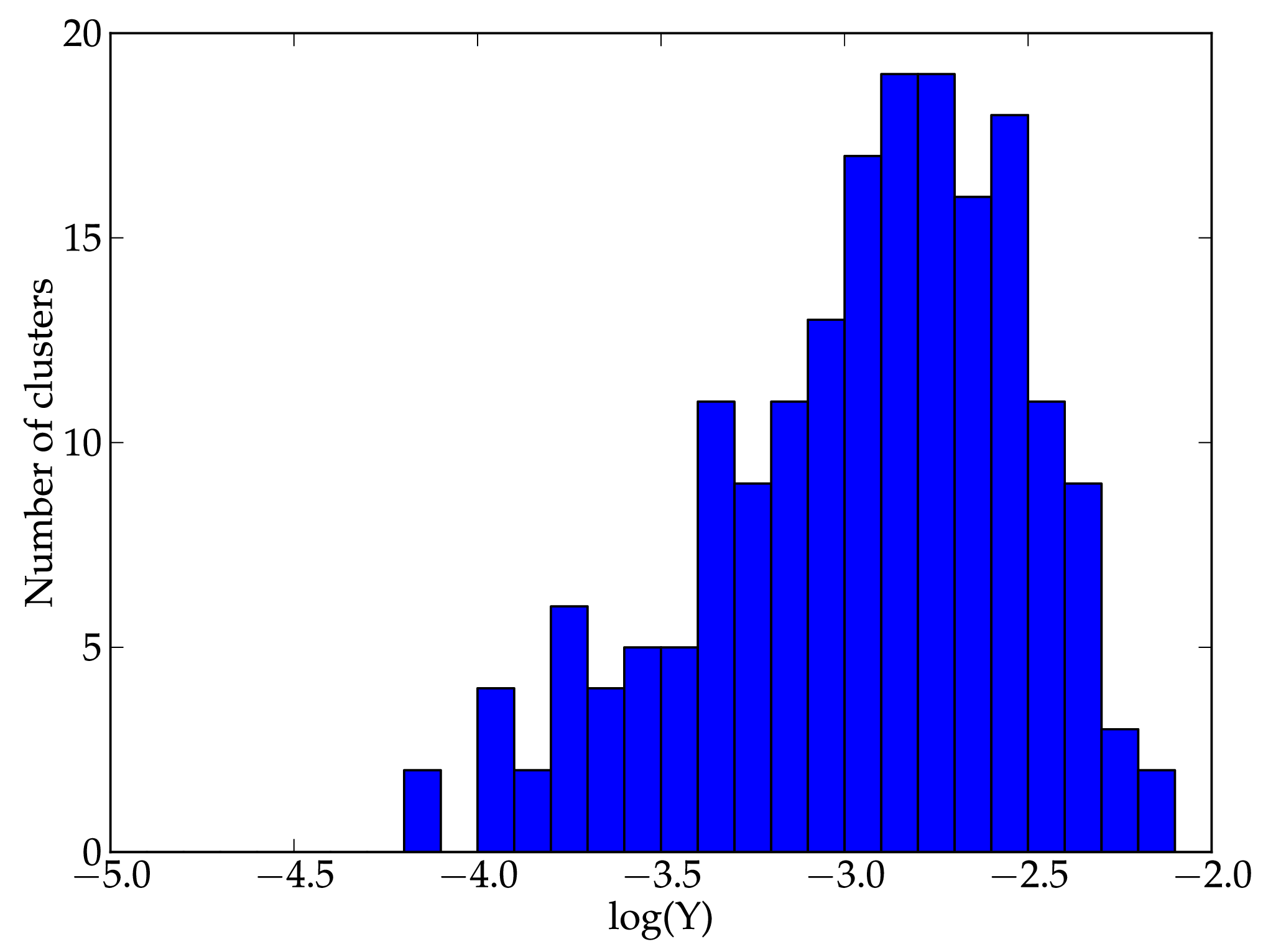}
\caption{ Histogram of the $Y$ values in logarithmic form obtained in one simulation, using the scaling law defined in Eq. \ref{Y-T} with $\alpha=0$ (\emph{top}) and $\alpha=-1$ (\emph{bottom}). An intrinsic dispersion of 10\% is assumed.}
\label{fig:Ydis}
\end{figure}

To build the largest possible sample, we made use of the X-ray Galaxy Cluster Database (BAX). We selected a sample containing all the identified clusters and groups for which temperatures are published in the literature with a relative measurement error of less than 15\%. This makes the sample far from being uniform but it serves the purpose of obtaining the largest sample of potentially observable clusters in upcoming SZ surveys with measured X-ray temperatures. Under these conditions, we were able to select a total of 438 clusters, with redshifts between 0.003 and 1. In Fig.\ref{fig:z_T} we plot the distribution of temperature versus redshift for the global sample, allowing for a better visualization of the sample measured quantities. In the same figure we show clusters that passed selection cuts described below in the paper. 

To perform our study, we started by generating "real'' temperatures for each cluster through a simple Gaussian sampling based on the measured mean values and uncertainties. These real temperatures were then used as a starting point to derive other observable quantities of the sample, based on the scaling laws presented above.  

One important feature when measuring the SZ flux is the angular size of the cluster in the sky. Large clusters will be fully resolved and an exact measurement of the $Y$ parameter up to the virial radius will require a good modeling of the cluster internal structure and pressure profile. On the other hand, smaller clusters will become nearly or completely unresolved and the virial measured integrated flux will thus become less model-dependent.

In Fig. \ref{fig:size_dist} we show the distribution of the angular radius for the clusters in our sample (for one realization), which presents a peak at about 8 arcmin and extends to values as high as 50 arcmin. These values were derived from Eq. \ref{MTv} together with the condition $M_V=4/3\pi \rho_m \delta_V R_V^3$ and the conversion to angular size using the cosmological angular distance. One can see that by introducing a significant evolution in the $M-T$ relation, the overall size distribution is still preserved. We decided to only select clusters with an apparent radial size of less than 10 arcmin, a value about twice higher than the expected angular beam size for the Planck SZ survey \footnote{http://www.rssd.esa.int/index.php?project=PLANCK\&page=perf\_top}. This choice was made because once we used the Planck resolution as a sample limit, the number of clusters would be drastically reduced to 30. Adopting this nearly resolved margin, and using 15 different realizations, the sample after this cut in angular size was composed of 199$\pm$1 clusters. This low variance reflects the good precision in the measured temperatures of the initial cluster sample.

We proceeded then to construct a "mock'' catalog of the observed $Y$ values. Because we are interested in the dependence of the $Y-T$ relation on redshift, which is modeled as a power law, we chose to work in logarithmic units for both $Y$ and $T$. In this way, the ``real'' $\log (Y)$ was obtained by taking the logarithm of Eq. \ref{MTv} and introducing the cosmological parameters in our fiducial model together with the ``real'' temperature of each clusters. Furthermore, we modeled an intrinsic dispersion of the scaling law of 10\% by Gaussian sampling around this first estimate of $\log (Y)$. This value corresponds to the typical result for the dispersion in $M-T_X$ or $M-Y$ obtained in numerical simulations (\cite{Kratsov2006}). Moreover, the recent analysis from the Planck collaboration (\cite{Plancka}) for low-redhift clusters seems to support a value of 12\% at the 1$\sigma$ level, consistent with 10\% at the 2$\sigma$ level.

To the values obtained after the introduction of intrinsic scatter we added random measurement errors based on the estimated flux detection threshold for the Planck SZ survey. Following \cite{melin} and the recent Planck data release (\cite{Planckc}), we considered the value for the threshold to be of $1.10^{-3}$ arcmin$^2$.  Moreover, we assumed the measurement errors to have a mean value corresponding to 20\% of the detection threshold, which corresponds to a signal-to-noise ratio of 5. After this procedure we obtained a simulation of the observed integrated flux parameters using our two fiducial models for $Y-T$ and $M-T$, with a distribution (for one realization) presented in Fig. 3. This distribution presents a peak around $10^{-3}$ arcmin$^2$ and extends up to $4.10^{-5}$ arcmin$^2$, thus confirming the importance of having a low detection threshold. In fact, by setting the detection threshold at $1.10^{-3}$ arcmin$^2$, the total number of detectable clusters is reduced to 79$\pm$4 (estimated from 15 realizations). If $\alpha=-1$, the sample size increases to 107$\pm$4.

\section{Constraining power of $Y-T$ observations}

In this section we explore the use of data on X-ray temperature and SZ flux to derive constraints on both the cosmological parameters and the parameters of the scaling laws in clusters. The "mock'' $Y$ data derived in the previous section were used to identify the constraining power and possible presence of bias when combining these two information sources.

We are thus interested in fitting the data with a scaling law (Eq.\ref{Y-T}), which in logarithmic form can be written as 

\begin{eqnarray}
\log Y =  K &+& \log \xi+ \frac{5}{2}\log T -\frac{3}{2} \left(1+\alpha\right)\log(1+z) \nonumber \\
          &-&\frac{1}{2}\log \Om - \log h - 2 \log d_a(\Om,\Ol,h)\;,
\label{logYlogT}
\end{eqnarray}
where $K$ is a constant factor accounting for terms like $f_{gas}$, physical constants and the normalization $A_{TM}$, while the variation in these parameters as well as the SZ contribution beyond the virial radius can be absorbed in our parameter $\xi$. The other terms are explicitly shown to better illustrate the linear dependence on the observables $\log T$, $\log (1+z)$. We considered two data sets, one constructed from a standard scaling law ($\alpha=0$), and a second data set comprising a strong non-standard evolution ($\alpha=-1$).     

\begin{table*}[t]
\caption{Table summarizing the mean posterior and 68\% confidence intervals of the constrained parameters.}
\label{Table1}
\centering
\begin{tabular}{c c c c c}
\hline
$\mathbf{Fiducial\; model}$	& $\alpha=0$	& $\alpha=-1 $	 &  $\alpha=0$     & $\alpha=-1$      \\
$\mathbf{Threshold\; (arcmin^2)}$	& $1.10^{-3} $	& $1.10^{-3} $	 &  $1.10^{-5}$     &  $1.10^{-5}$     \\
\hline 
$                     \xi$   &   $ 0.56\pm0.05(\pm0.04)$ & $  0.54\pm0.04(\pm0.03) $  &  $0.56\pm0.04(\pm0.04)$     &  $0.57\pm0.02(\pm0.03)$     \\       
$                     \alpha$&   $ -0.27\pm0.21(\pm0.18)$ & $ -1.29\pm0.15(\pm0.14) $  &  $-0.04\pm0.09(\pm0.10)$    &    $-0.99\pm0.09(\pm0.11)$    \\ 
     
\hline
\hline
$\mathbf{Fiducial\; model}$	& $\alpha=0$	& $\alpha=-1 $	 &  $\alpha=0$     & $\alpha=-1$      \\
$\mathbf{Threshold\; (arcmin^2)}$& $1.10^{-3} $	& $1.10^{-3} $	 &  $1.10^{-5}$     &  $1.10^{-5}$     \\
$\mathbf{\sigma_{disp}\; }$	& $ 0.078\pm0.040 $	& $ 0.068\pm0.049 $	 &  $ 0.113\pm0.018 $     &  $ 0.103\pm0.019 $     \\
\hline
$                     \xi$   &   $ 0.56\pm0.03(\pm0.05)$ & $  0.55\pm0.04(\pm0.04) $  &  $0.57\pm0.04(\pm0.03)$    &  $0.57\pm0.04(\pm0.03)$     \\       
$                     \alpha$&   $ -0.22\pm0.23(\pm0.18)$ & $ -1.29\pm0.17(\pm0.17) $  &  $-0.02\pm0.09(\pm0.10)$    &  $-0.98\pm0.09(\pm0.09)$    \\        
\hline
\end{tabular}
\tablefoot{Table summarizing the obtained constraints on scaling parameters and intrinsic dispersion. Four combinations of fiducial parameters and SZ detection limits are presented, using two approaches to the intrinsic dispersion of the scaling laws. The upper part of the table shows the results considering the same fiducial value as in the simulations ($\sigma_{disp}=0.1$) . The lower part shows the results when $\sigma_{disp}$ is derived from the simulated sample itself. For the parameters $\alpha$ and $\xi$ the first quoted errors correspond to the mean value of the 68\% confidence errors derived from each realization; the second quoted uncertainties represent the standard deviation of the mean central values from the fifteen realizations. The quoted uncertainties given for $\sigma_{disp}$ are the standard deviation derived from the fifteen realizations.}
\end{table*}

An important aspect to take into account when fitting this model is that both dependent and independent data present non negligible measurement errors. In this case, $Y$, $T$, and $z$ are measured with some degree of uncertainty and, although the cluster redshifts $z$ are usually measured with high precision, this is not the case for the temperature and SZ flux. It is necessary then to use a statistical estimator that takes this into account by constructing an effective $\chi^2$ using a total least squares regression technique. If $Y_i$, $T_i$, and $z_i$ are the measured values corresponding to the same cluster ($Y_i$ corresponding to our simulated values) and F($T_i$, $z_i$, $\xi$, $\alpha$) is the linear model as on the right hand side of Eq.\ref{logYlogT}, the total least-squares $\chi^2$ can be written as  

\begin{equation}
\chi_{eff}^2=\sum_i \frac{\left[\log Y_i-F(T_i,z_i,\xi,\alpha)\right]^2}{\sigma_{Y,i}^2+ \left(\frac{5}{2}\right)^2 \sigma_{T,i}^2+\sigma_{disp}^2}\;.
\label{chi2}
\end{equation}  
This expression assumes that the measurement errors $\sigma_{Y,i}$ and $\sigma_{T,i}$ are uncorrelated and neglects the contribution from redshift measurement errors ($\sigma_{z,i}$). Moreover, it takes into account the intrinsic scatter ($\sigma_{disp}$) around the scaling law.

The above estimator was used to constrain a flat $\Lambda$CDM model with six free parameters ($\Omega_{CDM}h^2$, $\Omega_{b}h^2$, h, $\tau$, n$_s$, A$_s$) plus the two free parameters of our scaling model $\xi$ and $\alpha$. We added a module to the public Markov Chain Monte Carlo code COSMOMC, and performed a joint constrain using CMB, LSS, SNIa and our SZ-Xray data. This combination can provide strong constraints on cosmological parameters required to reduce parameter degeneracies, which improves the constraints on cluster-related parameters.
  
As stated above, the nature of Monte-Carlo simulations is prone to an intrinsic variability when drawing different realizations. For this reason, we computed a total of 15 realizations for each considered fiducial model and detection threshold and used them to perform the MCMC chains. The mean values derived from the posterior distribution on the parameters of interest were then averaged to account for the intrinsic variability. For consistency, we compared the standard deviation for each parameter over the 15 realizations and verified that it was consistent with the typical 1$\sigma$ error for a single realization.


Another important aspect when using this kind of data is the estimation of the intrinsic dispersion $\sigma_{disp}$ to be accounted for in Eq. \ref{chi2}. Similarly to what is done with the SNIa Hubble diagram (\cite{Kowalski}; \cite{Hicken}), we estimated this quantity using a goodness-of-fit criterion and residual analysis. Although we knew by construction the real value for $\sigma_{disp}$, we adopted an iterative method to find the intrinsic dispersion using a maximum-likelihood criterion. In this case, the ``correct'' value should verify $\chi^2_{eff}\sim N-p$, with $N$ being the total number of clusters and $p$ the number of fitted parameters. To implement this, we ran MCMC chains up to 10000 points for a given data set and assuming an initial estimate of $\sigma_{disp}=0$. After having a reasonable estimate of the minimum $\chi^2$ value and the associated parameters, the residuals were fitted to obtain a first estimate of the intrinsic dispersion, which was then used to repeat the procedure. After a small number of iterations we let the MCMC run until full convergence, which produced a best-fit model with $\chi^2\approx N-p$. This allowed us to study the impact of the assumed intrinsic dispersion in the constrained parameters.   
 
 Table 1 summarizes the obtained results for the constrained parameters in the two fiducial models, assuming different values for the flux detection threshold. These values account for the averages and standard deviations obtained from the 15 realizations made for each case. We explicitly show the differences obtained when the correct intrinsic dispersion of 0.1 was assumed (upper part of the table) or when it was estimated using the method described above (lower part). In the later case, the values estimated for $\sigma_{disp}$ were also averaged over the 15 realizations and we provide the obtained mean and stdev values.

 The results show a noticeable dependence of the scaling parameter $\alpha$ on the assumed detection threshold. When the initial value of $1.10^{-3}$arcmin$^2$ is assumed, the marginalized mean values for the evolution parameters are slightly underestimated, with the fiducial values (used to construct the data set) lying outside the derived 68\% confidence intervals. This is most likely because the clusters with lower values of $Y$ (which are not detected at this threshold) are often those at higher redshifts or those presenting a significant deviation from the mean value. When a higher threshold value is considered, the absence of these undetected clusters will cause the slope of the $Y-z$ relation to be tilted towards lower values. A similar but smaller effect occurs with the normalization parameter.

\begin{figure}[t!]
\includegraphics[width=0.52\textwidth]{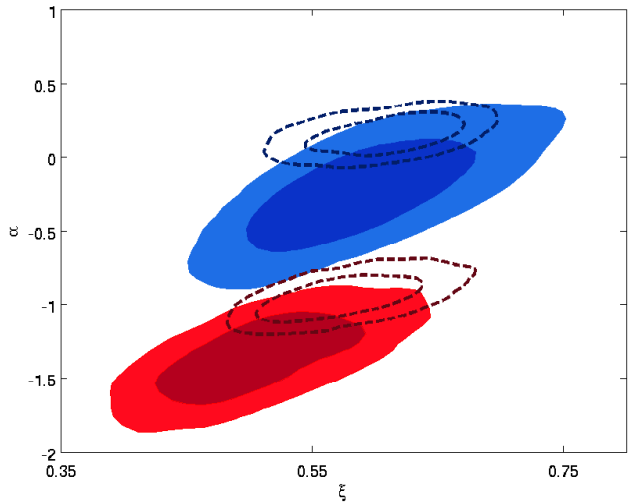}
\caption{ Two-dimensional marginalized constraints on the scaling model parameter, representing 68\% confidence regions for two-parameters. Blue (upper) contours were derived from data simulated with $\alpha=0$, and red (lower) contours using $\alpha=-1$. Filled contours assumed a $Y$ detection threshold of $1.10^{-3}$ arcmin$^2$, while dashed curves assumed this value to be $1.10^{-5}$ arcmin$^2$. }
\label{fig:contour}
\end{figure}      

To test this explanation, we used a lower value for the threshold in both cases to check if we could retrieve better estimates of the input fiducial parameters. Indeed, reducing the threshold to $1.10^{-5}$ arcmin$^2$ causes the number of detected clusters to increase significantly (detecting all the nearly unresolved clusters in our initial sample), as shown in Fig.1. This causes the constraints on the scaling parameters to become essentially unbiased and with a precision on $\alpha$ improved by a factor of 2 to 3, depending on the evolution scenario. In Fig. 4 we present the 2D confidence regions for four realizations with different fiducial values of $\alpha$ and detection threshold. The improvement in the constraints becomes evident when assuming a lower detection threshold. We note that the threshold does not have to be reduced to a value as low as $1.10^{-5}$ arcmin$^2$ to lead to a significant increase in the total number of detected clusters. Indeed, in Fig. \ref{fig:Ydis} one can see that only a small number of clusters remain undetected if one considers the threshold to be $1.10^{-4}$ arcmin$^2$. However, the few clusters that remain undetected have intermediate to high redshifts (between 0.5 and 1), which can be important to constrain evolution effects on scaling laws. Also, a lower threshold ensures better precision in individual cluster measurements.

Nevertheless, we computed MCMC chains considering a $1.10^{-4}$ arcmin$^2$ threshold and obtained a fairly intermediate result, i.e., larger uncertainties than those obtained with a $1.10^{-5}$ arcmin$^2$ threshold, but nearly consistent with the assumed fiducial values at 1$\sigma$ level.

 One can infer from Table 1 that the method used to estimate the intrinsic scatter of the scaling law ($\sigma_{disp}$) is able to retrieve the fiducial value within the 1$\sigma$ statistical variability of the Monte Carlo simulation for all studied cases. The decreasing of the detection threshold and the related inclusion of more clusters allows us to improve the accuracy with which the fiducial value is retrieved. We also note that the obtained constraints are highly stable, independent of the assumed method to account for the intrinsic dispersion $\sigma_{disp}$. 

\section{Conclusions}

We performed a study of the capacity of using measurements on the SZ integrated flux to constrain a possible non-standard evolution of the scaling relations of galaxy clusters. Based on simulated data derived from measured X-ray temperatures, we showed that a sample of 199$\pm$1 clusters with measured temperatures (comprising 79$\pm$4  detectable and nearly unresolved clusters within Planck's SZ observational "settings'') can indeed provide meaningful information on this subject, and distinguish between a strong non-standard evolution and the standard gravitationally driven scaling law. 

We showed that even if some intrinsic dispersion is present in the actual scaling law for $Y-T$, its amplitude can be estimated and extracted from the data with an acceptable  precision, using a goodness-of-fit analysis. More important is that this intrinsic dispersion does not affect the parameter estimates, allowing for a possible distinction between standard scaling models and strongly evolving ones.  

Using a conservative value for the SZ flux measurement threshold ($1.10^{-3}$ arcmin$^2$), the relative uncertainty obtained on the evolution index $\alpha$ is significant, because of the low number of unresolved detectable clusters. Furthermore, we showed that by imposing this relatively high flux limit, the obtained constraints on this parameter present a significant bias that needs to be taken into account. This implies that although the Planck SZ survey can provide an important distinction between non-evolving and strongly evolving $Y-T$ relation, its results can be affected by a non-negligible systematic error. Precision temperature measurements through an X-ray follow-up of newly SZ discovered clusters could in principle help to reduce parameter uncertainties, and we showed that lowering the $Y$ detection flux limit allows one to obtain unbiased estimates for the evolution parameters.  

After the submission of this paper, the Planck collaboration has released its early results in a set of papers, with several ones focusing on the SZ detection and measurements of cluster properties. (\cite{Plancka}, \cite{Planckb}, \cite{Planckc}). Although these comprise an analysis of the scaling relations at low redshift, the method used to treat the Planck data is based on the assumption of a universal profile and scaling law with parameters derived from X-ray data alone.  In these analyses, the measurements are restricted to the central part of clusters. Our work shows that it should be possible to address the possible non-standard history of the thermal content of clusters, provided that the bias induced in $Y$ by selection effects is properly taken into account. 

In the wake of a new era for SZ cluster studies, the use of SZ flux measurements together with X-ray data can undoubtedly shed some light on the evolution of the properties of these massive objects, which can result in important consequences for cosmological applications.

\acknowledgements{
We thank Yves Zolnierowski for useful discussions. We also thank the anonymous referee for useful comments and suggestions. This research has made use of the X-Rays Galaxy Clusters Database (BAX), which is operated by the Institut de Recherche en Astrophysique et Planetologie (IRAP). Luis Ferramacho acknowledges financial support from Region Midi-Pyr\'en\'ees.}

\end{document}